\documentclass[finalprint,3p]{elsarticle}
\usepackage{amsmath}
\usepackage{lineno,hyperref}

\modulolinenumbers[5]

\journal{Journal of \LaTeX\ Templates}

\hypersetup{colorlinks=true,linkcolor=blue,citecolor=blue,filecolor=blue,urlcolor=blue,breaklinks=true}

\begin{document}

\begin{frontmatter}

\title{Quasi-exactly-solvable confining solutions for spin-1 and spin-0
bosons in (1+1)-dimensions with a scalar linear potential}

\author[mymainaddress]{Luis B. Castro\corref{mycorrespondingauthor}}
\cortext[mycorrespondingauthor]{Corresponding author}
\ead{lrb.castro@ufma.br}

\author[mysecondaryaddress]{Antonio S. de Castro}
\ead{castro@pq.cnpq.br}

\address[mymainaddress]{Departamento de F\'{\i}sica, Universidade Federal do Maranh\~{a}o, Campus Universit\'{a}rio do Bacanga,\\ 65080-805, S\~{a}o Lu\'{\i}s, MA, Brazil.}
\address[mysecondaryaddress]{Departamento de F\'{\i}sica e Qu\'{\i}mica, Universidade Estadual Paulista, Campus de Guaratin\-gue\-t\'{a},\\ 12516-410, Guaratinguet\'{a}, SP, Brazil.}

\begin{abstract}
We point out a misleading treatment in the recent literature regarding
confining solutions for a scalar potential in the context of the
Duffin-Kemmer-Petiau theory. We further present the proper bound-state
solutions in terms of the generalized Laguerre polynomials and show that the
eigenvalues and eigenfunctions depend on the solutions of algebraic
equations involving the potential parameter and the quantum number.
\end{abstract}


\end{frontmatter}

\linenumbers

\section{Introduction}

The Duffin-Kemmer-Petiau (DKP) formalism \cite{Petiau1936,Kemmer1938,PR54:1114:1938,Kemmer1939} describes spin-0 and
spin-1 bosons and has been used to analyze relativistic interactions of
spin-0 and spin-1 hadrons with nuclei as an alternative to their
conventional second-order Klein-Gordon (KG) and Proca counterparts (see,
e.g. \cite{JPA43:055306:2010} for a comprehensive list of references). The DKP formalism
enjoys a richness of couplings not capable of being expressed in the KG and
Proca theories \cite{PRD15:1518:1977,JPA12:665:1979}. Although the formalisms are equivalent in the
case of minimally coupled vector interactions \cite{PLA244:329:1998,PLA268:165:2000,PRA90:022101:2014}, the DKP
formalism opens new horizons as far as it allows other kinds of couplings
which are not possible in the KG and Proca theories.

The scalar interaction refers to a kind of coupling that behaves like a
scalar (invariant) under a Lorentz transformation. Though the scalar interaction
finds many of their applications in nuclear and particle physics, it could also simulate an effective mass in solid state physics. Due to weak potentials, relativistic effects are considered to be small in solid state physics, but the relativistic wave equations can give relativistic corrections to the results obtained from the nonrelativistic wave equation, therefore the relativistic extension of this problem is also of interest and remains unexplored. The scalar interaction in the context of the DKP
theory has been reported in the literature for a smooth step potential \cite%
{IJTP46:2105:2007}, step potential \cite{IJTP49:10:2010}, Coulomb potential \cite{PRC84:064003:2011} and
linear potential \cite{PS87:065003:2013}. In Ref.~\cite{PS87:065003:2013}, the authors examined the
time-independent DKP equation in a (1+1)-dimension with the scalar linear
potential and claimed to have obtained exact conditions selecting the
eigenvalues. In that paper, the authors misidentified the correct asymptotic
behavior of Kemmer's function, kept out Kummer's function and used Tricomi's
function as a particular solution.

The purpose of this paper is to review the DKP equation in the presence of a
scalar linear potential for spin-1 and spin-0 bosons in (1+1)-dimensions.
Following the appropriate \textit{modus operandi}, we show that the problem
is exactly solvable on the whole line for a restrict class of potential
parameters and quantum numbers. In this circumstance, the eigenenergies are
solutions of algebraic equations and the eigenfunctions are expressed in
terms of the generalized Laguerre polynomials.

\section{The Duffin-Kemmer-Petiau equation}

The DKP equation for a free boson is given by \cite{Kemmer1939} (with units in
which $\hbar =c=1$)%
\begin{equation}
\left( i\beta ^{\mu }\partial _{\mu }-m\right) \Psi =0  \label{dkp}
\end{equation}
\noindent where the matrices $\beta ^{\mu }$\ satisfy the algebra%
\begin{equation}
\beta ^{\mu }\beta ^{\nu }\beta ^{\lambda }+\beta ^{\lambda }\beta ^{\nu
}\beta ^{\mu }=g^{\mu \nu }\beta ^{\lambda }+g^{\lambda \nu }\beta ^{\mu }
\label{beta}
\end{equation}
\noindent and the metric tensor is $g^{\mu \nu }=\,$diag$\,(1,-1,-1,-1)$.
The algebra expressed by (\ref{beta}) generates a set of 126 independent
matrices whose irreducible representations are a trivial representation, a
five-dimensional representation describing the spin-0 particles and a
ten-dimensional representation associated to spin-1 particles. The
second-order KG and Proca equations are obtained when one selects the spin-0
and spin-1 sectors of the DKP theory. A well-known conserved four-current is
given by
\begin{equation}
J^{\mu }=\frac{1}{2}\bar{\Psi}\beta ^{\mu }\Psi   \label{corrente}
\end{equation}
\noindent where the adjoint spinor $\bar{\Psi}=\Psi ^{\dagger }\eta ^{0}$,
with $\eta ^{0}=2\beta ^{0}\beta ^{0}-1$ in such a way that $\left( \eta
^{0}\beta ^{\mu }\right) ^{\dagger }=\eta ^{0}\beta ^{\mu }$ (the matrices $%
\beta ^{\mu }$ are Hermitian with respect to $\eta ^{0}$). Despite the
similarity to the Dirac equation, the DKP equation involves singular
matrices, the time component of $J^{\mu }$ is not positive definite but it
may be interpreted as a charge density. The factor $1/2$ multiplying $\bar{%
\Psi}\beta ^{\mu }\Psi $, of no importance regarding the conservation law,
is in order to hand over a charge density conformable to that one used in
the KG theory and its nonrelativistic limit \cite{JPA43:055306:2010}. Then the
normalization condition
\begin{equation}\label{nor}
    \int d\tau \,J^{0}=\pm 1
\end{equation}
\noindent can be expressed as%
\begin{equation}
\int d\tau \,\bar{\Psi}\beta ^{0}\Psi =\pm 2  \label{norm}
\end{equation}
\noindent where the plus (minus) sign must be used for a positive (negative)
charge.

\subsection{Interaction in the DKP equation}

With the introduction of interactions, the DKP equation can be written as%
\begin{equation}
\left( i\beta ^{\mu }\partial _{\mu }-m-U\right) \Psi =0  \label{dkp2}
\end{equation}
\noindent where the more general potential matrix $U$ is written in terms of
25 (100) linearly independent matrices pertinent to five (ten)-dimensional
irreducible representation associated to the scalar (vector) sector. In the
presence of interaction, $J^{\mu }$ satisfies the equation
\begin{equation}
\partial _{\mu }J^{\mu }+\frac{i}{2}\,\bar{\Psi}\left( U-\eta ^{0}U^{\dagger
}\eta ^{0}\right) \Psi =0  \label{corrent2}
\end{equation}
\noindent Thus, if $U$ is Hermitian with respect to $\eta ^{0}$ then
four-current will be conserved. The potential matrix $U$ can be written in
terms of well-defined Lorentz structures. For the spin-zero sector there are
two scalar, two vector and two tensor terms \cite{PRD15:1518:1977}, whereas for the
spin-one sector there are two scalar, two vector, a pseudoscalar, two
pseudovector and eight tensor terms \cite{JPA12:665:1979}. The tensor terms have been
avoided in applications because they furnish noncausal effects \cite{PRD15:1518:1977,JPA12:665:1979}. The condition (\ref{corrent2}) has been used to point out a
misleading treatment in the recent literature regarding analytical solutions
for nonminimal vector interactions \cite{CJP87:857:2009,CJP87:1185:2009,JPA45:075302:2012,ADHEP2014:784072:2014}.

\subsection{Scalar coupling in the DKP equation}

Considering only scalar interaction, the DKP equation can be written as
\begin{equation}
\left( i\beta ^{\mu }\partial _{\mu }-m-S\right) \Psi =0  \label{dkp2}
\end{equation}
\noindent with $S$ denoting the scalar potential function.

For the case of spin-0, we use the representation for the $\beta ^{\mu }$\
matrices given by \cite{JPG19:87:1993}%
\begin{equation}
\beta ^{0}=%
\begin{pmatrix}
\theta  & \overline{0} \\
\overline{0}^{T} & \mathbf{0}%
\end{pmatrix}%
,\quad \beta ^{i}=%
\begin{pmatrix}
\widetilde{0} & \rho _{i} \\
-\rho _{i}^{T} & \mathbf{0}%
\end{pmatrix}%
,\quad i=1,2,3  \label{rep}
\end{equation}%
\noindent where%
\begin{eqnarray}
\ \theta  &=&%
\begin{pmatrix}
0 & 1 \\
1 & 0%
\end{pmatrix}%
,\quad \rho _{1}=%
\begin{pmatrix}
-1 & 0 & 0 \\
0 & 0 & 0%
\end{pmatrix}
\notag \\
&&  \label{rep2} \\
\rho _{2} &=&%
\begin{pmatrix}
0 & -1 & 0 \\
0 & 0 & 0%
\end{pmatrix}%
,\quad \rho _{3}=%
\begin{pmatrix}
0 & 0 & -1 \\
0 & 0 & 0%
\end{pmatrix}
\notag
\end{eqnarray}%
\noindent $\overline{0}$, $\widetilde{0}$ and $\mathbf{0}$ are 2$\times $3, 2%
$\times $2 \ and 3$\times $3 zero matrices, respectively, while the
superscript T designates matrix transposition. The five-component spinor can
be written as $\Psi ^{T}=\left( \Psi _{1},...,\Psi _{5}\right) $ in such a
way that the DKP equation for a boson constrained to move along the $x$-axis
decomposes into
\begin{equation*}
\partial _{0}\Psi _{1}=-i\left( m+S\right) \Psi _{2},\quad \partial _{1}\Psi
_{1}=-i\left( m+S\right) \Psi _{3}
\end{equation*}%
\begin{equation}
\partial _{0}\Psi _{2}-\partial _{1}\Psi _{3}=-i\left( m+S\right) \Psi _{1}
\label{DKP3}
\end{equation}%
\begin{equation*}
\Psi _{4}=\Psi _{5}=0
\end{equation*}
\noindent and $J^{\mu }$ can be written as%
\begin{equation}
J^{0}=\text{Re}\left( \Psi _{2}^{\ast }\Psi _{1}\right) ,\quad J^{1}=-\text{%
Re}\left( \Psi _{3}^{\ast }\Psi _{1}\right) ,\quad J^{2}=J^{3}=0
\label{corrente3}
\end{equation}

For the case of spin-1, the $\beta ^{\mu }$\ matrices are \cite{JMP35:4517:1994}%
\begin{equation}
\beta ^{0}=%
\begin{pmatrix}
0 & \overline{0} & \overline{0} & \overline{0} \\
\overline{0}^{T} & \mathbf{0} & \mathbf{I} & \mathbf{0} \\
\overline{0}^{T} & \mathbf{I} & \mathbf{0} & \mathbf{0} \\
\overline{0}^{T} & \mathbf{0} & \mathbf{0} & \mathbf{0}%
\end{pmatrix}%
,\;\beta ^{i}=%
\begin{pmatrix}
0 & \overline{0} & e_{i} & \overline{0} \\
\overline{0}^{T} & \mathbf{0} & \mathbf{0} & -is_{i} \\
-e_{i}^{T} & \mathbf{0} & \mathbf{0} & \mathbf{0} \\
\overline{0}^{T} & -is_{i} & \mathbf{0} & \mathbf{0}%
\end{pmatrix}
\label{betaspin1}
\end{equation}%
\noindent where $s_{i}$ are the 3$\times $3 spin-1 matrices $\left(
s_{i}\right) _{jk}=-i\varepsilon _{ijk}$, $e_{i}$ are the 1$\times $3
matrices $\left( e_{i}\right) _{1j}=\delta _{ij}$ and $\overline{0}=%
\begin{pmatrix}
0 & 0 & 0%
\end{pmatrix}%
$, while\textbf{\ }$\mathbf{I}$ and $\mathbf{0}$\textbf{\ }designate the 3$%
\times $3 unit and zero matrices, respectively. The spinor $\Psi ^{T}=\left(
\Psi _{1},...,\Psi _{10}\right) $ can be partitioned as%
\begin{equation}
\Psi _{I}^{T}=\left( \Psi _{3},\Psi _{4},\Psi _{5}\right) ,\quad \Psi
_{II}^{T}=\left( \Psi _{6},\Psi _{7},\Psi _{2}\right) ,\quad \Psi
_{III}^{T}=\left( \Psi _{10},-\Psi _{9},\Psi _{1}\right)   \label{spinor}
\end{equation}%
\noindent so that the one-dimensional DKP equation can be expressed in the form
\begin{equation*}
\partial _{0}\Psi _{I}=-i\left( m+S\right) \Psi _{II},\quad \partial
_{1}\Psi _{I}=-i\left( m+S\right) \Psi _{III}
\end{equation*}%
\begin{equation}
\partial _{0}\Psi _{II}-\partial _{1}\Psi _{III}=-i\left( m+S\right) \Psi
_{I}\label{DKp3}
\end{equation}%
\begin{equation*}
\Psi _{8}=0
\end{equation*}%
\noindent In addition, expressed in terms of (\ref{spinor}) the current can be written
as
\begin{equation}
J^{0}=\text{Re}\left( \Psi _{II}^{\dagger }\Psi _{I}\right) ,\quad
J^{1}=-\text{Re}\left( \Psi _{III}^{\dagger }\Psi _{I}\right) ,\quad
J^{2}=J^{3}=0  \label{C1}
\end{equation}%
\noindent Comparison of (\ref{DKP3}) with (\ref{DKp3}) evidences that the spinors $%
\Psi _{I}$, $\Psi _{II}$ and $\Psi _{III}$ behave like the spinor components
$\Psi _{1}$, $\Psi _{2}$ and $\Psi _{3}$, respectively, from the spin-0
sector of the DKP theory. More than this, comparison of (\ref{corrente3})
with (\ref{C1}) places on view that the spin-1 sector of the DKP theory
looks formally like the spin-0 sector \cite{IJTP49:10:2010}.

For a time-independent scalar potential, one can write $\Psi (x,t)=\psi
(x)\exp (-iEt)$. With the abbreviations $\phi =\psi _{1}$ ($\psi _{I}$), $%
\phi _{2}=\psi _{2}$ ($\psi _{II}$) and $\phi _{3}=\psi _{3}$ ($\psi _{III}$%
) the time-independent DKP equation for the spin-0 (spin-1) sector of the
DKP theory splits into%
\begin{equation}
\frac{d}{dx}\left( \frac{1}{m+S}\frac{d\phi }{dx}\right) +\frac{E^{2}-\left(
m+S\right) ^{2}}{m+S}\phi=0  \label{eq1}
\end{equation}%
\begin{equation}
\phi _{2}=\frac{E}{m+S}\phi
\end{equation}%
\begin{equation}
\phi _{3}=\frac{i}{m+S}\frac{d\phi }{dx}
\end{equation}%
\noindent with%
\begin{equation}\label{j0}
J^{0}=\frac{E}{m+S}|\phi |^{2},\quad J^{1}=\frac{1}{m+S}\textrm{Im}\left( \phi
^{\dag }\frac{d\phi }{dx}\right)
\end{equation}

\subsection{ Linear potential}

In Ref. \cite{PS87:065003:2013}, the authors used%
\begin{equation}
S\left( x\right) =\lambda |x|,\quad \lambda >0  \label{eq7}
\end{equation}%
\noindent and the changes%
\begin{equation}
z=1+\frac{\lambda }{m}|x|,\qquad \phi \left( z\right) =z^{1+\beta
}e^{-z^{2}/2g}f\left( z\right)   \label{eq11}
\end{equation}%
\noindent with the definitions%
\begin{equation}
g=\frac{\lambda }{m^{2}},\qquad \varepsilon =\frac{E}{gm},\qquad \beta ^{2}=1
\label{eq12}
\end{equation}%
\noindent With the additional change of variable%
\begin{equation}
t=\frac{z^{2}}{g}
\end{equation}%
\noindent they finally arrived at%
\begin{equation}
t\frac{d^{2}f\left( t\right) }{dt^{2}}+\left( \beta +1-t\right) \frac{%
df\left( t\right) }{dt}+\left( \frac{g\varepsilon ^{2}}{4}-\frac{\beta +1}{2}%
\right) f\left( t\right) =0  \label{eq13}
\end{equation}%
\noindent and presented the solution
\begin{equation}
f\left( t\right) =A\,M(a,b,t)+B\,U\left( a,b,t\right)   \label{sol1}
\end{equation}%
\noindent where $A$ and $B$ are arbitrary constants, $M(a,b,t)$ and $U(a,b,t)$ are
Kummer's and Tricomi's functions respectively, and%
\begin{equation}
a=\frac{\beta +1}{2}-\frac{g\varepsilon ^{2}}{4},\qquad b=\beta +1
\end{equation}%
\noindent All of their remaining analysis rested on identifying Eq. (\ref{sol1}) with
the general solution of the confluent hypergeometric equation. Furthermore,
they assumed that the asymptotic behaviour of Kummer's function is given by $%
\,e^{t}\,t^{a-b}$, kept out Kummer's function and used Tricomi's function as
a particular solution.

\section{Solutions on the half line}

For solutions of the confluent hypergeometric equation and their properties
we refer to Abramowitz and Stegun \cite{ABRAMOWITZ1965}. Kummer's function is expressed
as
\begin{equation}
M(a,b,w)=\frac{\Gamma \left( b\right) }{\Gamma \left( a\right) }%
\sum_{j=0}^{\infty }\frac{\Gamma \left( a+j\right) }{\Gamma \left(
b+j\right) }\,\frac{w^{j}}{j!}=1+\frac{a}{b}\frac{w}{1!}+\frac{a\left(
a+1\right) }{b\left( b+1\right) }\frac{w^{2}}{2!}+\cdots
\end{equation}%
\noindent where $\Gamma \left( w\right) $ is the gamma function with simple
poles at $w=0,-1,-2,-3,\ldots $ and the defining property $\Gamma \left(
w+1\right) =w\Gamma \left( w\right) $. On the other hand, Tricomi's function
is expressed in terms of Kummer's function as%
\begin{equation}
U(a,b,w)=\frac{\pi }{\sin \pi b}\left[ \frac{M(a,b,w)}{\Gamma \left(
1+a-b\right) \Gamma \left( b\right) }-w^{1-b}\frac{M(1+a-b,2-b,w)}{\Gamma
\left( a\right) \Gamma \left( 2-b\right) }\right]  \label{tri}
\end{equation}%
\noindent Notice that $M\left( a,b,w\right) $ corresponds to $F\left(
a,b,w\right) $ in Ref.~\cite{PS87:065003:2013}. Notice also that the identity $\Gamma
\left( w\right) \Gamma \left( 1-w\right) =\pi /\sin \pi w$ carries Eq.~(\ref%
{tri}) on exactly the same form as Eq.~(28) in Ref.~\cite{PS87:065003:2013}. The Wronskian
of $M(a,b,w)$ and $U(a,b,w)$ is given by%
\begin{equation}
W\left( M,U\right) =-\frac{\Gamma \left( b\right) }{\Gamma \left( a\right) }%
\,w^{-b}e^{w}  \label{w}
\end{equation}%
\noindent so that these functions are linearly independent solutions of the
confluent hypergeometric equation only if $a\neq -n$, where $n$ is a
nonnegative integer, and $b\neq -1,-2,-3,\ldots $\ Furthermore, these
functions present asymptotic behaviours as $|w|\rightarrow \infty $ dictated
by%
\begin{eqnarray}
&&M(a,b,w)\,_{\simeq }\,\frac{\Gamma \left( b\right) }{\Gamma \left(
b-a\right) }\,e^{-i\pi a}\,w^{-a}+\frac{\Gamma \left( b\right) }{\Gamma
\left( a\right) }\,e^{w}\,w^{a-b}  \notag \\
&&  \label{asym} \\
&&\quad U(a,b,w)_{\simeq }w^{-a}  \notag
\end{eqnarray}

It is true that the presence of $e^{w}$ in the asymptotic behaviour of $%
M(a,b,w)$ perverts the normalizability of $\phi \left( w\right) $.
Nevertheless, this unfavorable behaviour can be remedied by demanding $a=-n$
and $b\neq -\tilde{n}$, where $\tilde{n}$ is also a nonnegative integer. In
this case, one has to consider $b=2$ ($\beta =+1$). Therefore, the
asymptotic behaviour of Kummer's function is proportional to $w^{n}$ and
Tricomi's function becomes proportional to Kummer's function.

As a matter of fact, $M(-n,b,w)$ with $b>0$ and $w\in \lbrack 0,\infty )$ is
proportional to the generalized Laguerre polynomial $L_{n}^{\left(
b-1\right) }(w)$, a polynomial of degree $n$ with $n$ distinct positive
zeros in the range $[0,\infty )$. The requirement $a=-n$ and $b=2$ implies
into the solution on the half line%
\begin{eqnarray}
E_{n} &=&\pm \frac{2m}{\sqrt{\zeta }}\sqrt{n+1}  \notag \\
&&  \label{Eq10} \\
\phi _{n}\left( |x|\right)  &=&N_{n}te^{-t/2}L_{n}^{\left( 1\right) }\left(
t\right)   \notag
\end{eqnarray}%
\noindent where $t$ can be written as%
\begin{equation}
t=\zeta \left( 1+\,\frac{|x|}{\zeta \lambda _{C}}\right) ^{2}
\end{equation}%
\noindent $N_{n}$ is a normalization constant, $\zeta =1/g>0$ and $\lambda
_{C}=1/m$. It is useful to list the first few generalized Laguerre
polynomials $L_{n}^{\left( 1\right) }(w)$ as standardized in Ref. \cite{ABRAMOWITZ1965}:%
\begin{eqnarray}
L_{0}^{\left( 1\right) }(w) &=&1  \notag \\
L_{1}^{\left( 1\right) }(w) &=&-w+2 \\
L_{2}^{\left( 1\right) }(w) &=&w^{2}/2-3w+3  \notag
\end{eqnarray}%
\noindent and in general we have%
\begin{equation}
L_{n}^{\left( 1\right) }\left( w\right) =\sum_{j=0}^{n}\frac{\Gamma \left(
n+2\right) \left( -w\right) ^{j}}{\Gamma \left( j+2\right) j!\left(
n-j\right) !}  \label{her}
\end{equation}%
\noindent The following differential property%
\begin{equation}
w\frac{dL_{n}^{\left( 1\right) }\left( w\right) }{dw}=nL_{n}^{\left(
1\right) }\left( w\right) -\left( n+1\right) L_{n-1}^{\left( 1\right)
}\left( w\right)   \label{re}
\end{equation}%
\noindent allows us to expand the derivative of a generalized Laguerre
polynomial in terms of other generalized Laguerre polynomials with the same
superscripts. $L_{-1}^{\left( 1\right) }\left( w\right) $ is to be
interpreted as zero.

\section{Solutions on the whole line}

Following the ideas of the preceding section, we proceed now to find the
eigenfunctions on the whole line. Because jump discontinuities of \ $\phi
_{n}\left( x\right) $ and $d\phi _{n}\left( x\right) /dx$ would imply in the
presence of Dirac delta functions and their first derivatives in Eq.~(\ref%
{eq1}), respectively, lawful symmetric and antisymmetric extensions of $\phi
\left( |x|\right) $ given on the half line to the whole line are possible
only if $\phi _{n}\left( x\right) $ and $d\phi _{n}\left( x\right) /dx$ are
continuous at the origin. The continuity requirement implies that $\left.
d\phi _{n}\left( |x|\right) /dx\right\vert _{x=0_{+}}$ vanishes for an even
function, and $\left. \phi _{n}\left( |x|\right) \right\vert _{x=0_{+}}$
vanishes for an odd function. One has%
\begin{equation}
\left. \phi _{n}\left( |x|\right) \right\vert _{x=0_{+}}=N_{n}\zeta
e^{-\zeta /2}L_{n}^{\left( 1\right) }\left( \zeta \right)
\end{equation}%
\noindent and, with the aid of Eq.~(\ref{re}), one finds%
\begin{equation}
\left. \frac{d\phi _{n}\left( |x|\right) }{dx}\right\vert _{x=0_{+}}=\frac{%
2N_{n}e^{-\zeta /2}\left( n+1\right) }{\lambda _{C}}\left\{ \left[ 1-\frac{%
\zeta }{2\left( n+1\right) }\right] L_{n}^{\left( 1\right) }\left( \zeta
\right) -L_{n-1}^{\left( 1\right) }\left( \zeta \right) \right\}
\label{diff}
\end{equation}

When $L_{n}^{\left( 1\right) }\left( \zeta \right) \neq 0$ one finds $\left.
\phi _{n}\left( |x|\right) \right\vert _{x=0_{+}}\neq 0$. In this case, one
obtains from Eq. (\ref{diff}) that $\left. d\phi _{n}\left( |x|\right)
/dx\right\vert _{x=0_{+}}$ vanishes if $\zeta $ and $n$ satisfy the
following ($n+1$)-degree algebraic equation in $\zeta $:%
\begin{equation}
1-\frac{\zeta }{2\left( n+1\right) }=\frac{L_{n-1}^{\left( 1\right) }\left(
\zeta \right) }{L_{n}^{\left( 1\right) }\left( \zeta \right) }  \label{res}
\end{equation}%
\noindent Thus, the solution given by Eq.~(\ref{Eq10}) is acceptable on the
whole line as an even-parity function only if the potential parameter
satisfies the constraint relation expressed by Eq.~(\ref{res}). In
particular $\zeta =2$ with $n=0$, and $\zeta =3\pm \sqrt{5}$ with $n=1$.

On the other hand, the odd-parity solutions are related to the zeros of the
generalized Laguerre polynomial: $L_{n}^{\left( 1\right) }\left( \zeta
\right) =0$. This is a $n$-degree algebraic equation in $\zeta $ depending
on $n$. There are $n$ positive roots for a given $n$. One finds no solution
with $n=0$. Nevertheless, one finds $\zeta =2$ with $n=1$, and $\zeta =3\pm
\sqrt{3}$ with $n=2$, for example.

\begin{figure}[th]
\begin{center}
\includegraphics[width=8cm]{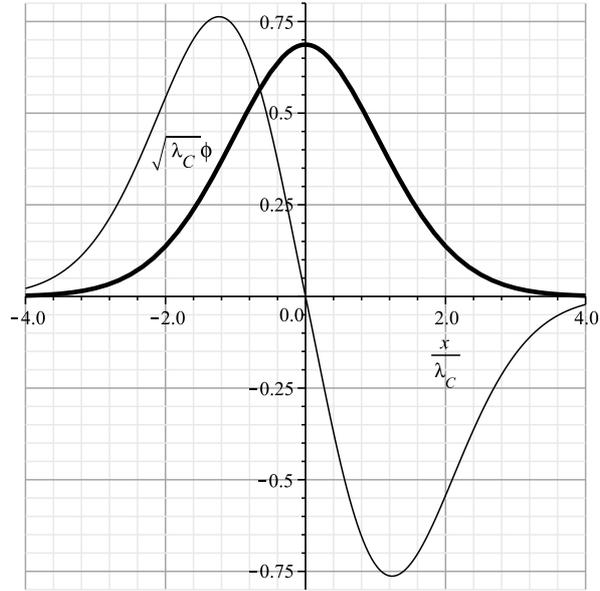}
\end{center}
\par
\vspace*{-0.1cm}
\caption{Normalized $\protect\sqrt{\protect\lambda _{C}}\,\protect\phi $ as
a function of $x/\protect\lambda _{C}$ for $\protect\zeta =2$. Heavy line
for $n=0$, and light line for $n=1$. The eigenfunctions are scaled in order
to be dimensionless.}
\label{fig1}
\end{figure}

Using (\ref{nor}) and (\ref{j0}), the normalization constant can be written as
\begin{equation}\label{constnor}
    N_{n}= \sqrt{\frac{\lambda}{\delta|E_{n}|}}\,.
\end{equation}
\noindent where $\delta=\int_{1/g}^{\infty}t\mathrm{e}^{-t}|L_{n}^{(1)}(t)|^{2}dt$. In Figure \ref{fig1}, we illustrate the results for $\zeta =2$.
\noindent Although $n$ is equal to the number of nodes of $\phi _{n}$ in
Figure \ref{fig1}, one should not expect this relation in a systematic way
because $t$ in the second line of Eq.~(\ref{Eq10}) is restricted to the
interval $[1/g,\infty )$. By way of addition, one may expect a two-fold
degeneracy, with even and odd eigenfunctions for the same potential
parameter and quantum number when both $\phi _{n}\left( x\right) $ and $%
d\phi _{n}\left( x\right) /dx$ vanish at the origin in such a way that $%
\zeta $ and $n$ satisfy the following set of equations: $L_{n}^{\left(
1\right) }\left( \zeta \right) =0$ and $L_{n-1}^{\left( 1\right) }\left(
\zeta \right) =0$. Nevertheless, no numerical solution is found for this
system of equations for $n\leq 100$, showing that the zeros of $%
L_{n}^{\left( 1\right) }\left( w\right) $ are different from the zeros of $%
L_{n-1}^{\left( 1\right) }\left( w\right) $. After all, this absence of
degeneracy is plausible in view of the so-called nondegeneracy theorem for
bound states in one-dimensional nonsingular potentials \cite{LANDAU1958}.

\section{Final remarks}

As commented in the Introduction, the authors of Ref.~\cite{PS87:065003:2013}
misidentified the correct asymptotic behaviour of Kummer's function, kept
out Kummer's function and used Tricomi's function as a particular solution.
These facts should be enough to nullify the candidature of the set of
solutions presented in Ref.~\cite{PS87:065003:2013} as bona fide solutions. Surprisingly,
that set of solutions is licit because $U(a,0,t)$ presents a good behaviour
at the neighbourhood of $1/g$ as well as a good asymptotic behaviour.
Nevertheless, the solutions have to be found by numerical methods.

We analyzed in detail the solutions of Eq.~(\ref{eq1}) with the linear
potential by given careful consideration to asymptotic behaviour of Kummer's
function. In that process, we have shown that the Sturm-Liouville problem
has been transmuted in a simpler problem of solving algebraic equations for
the eigenvalues and that the eigenfunctions are expressed in terms of the
generalized Laguerre polynomials. The quantization condition comes into
sight already for the problem defined on the half line and the extensions
for the whole line imply into extra algebraic equations constraining the
potential parameter and the quantum number. In general, one finds different
potential parameters for different quantum numbers.

\section*{References}

\bibliography{mybibfile2}

\end{document}